\documentstyle[camera,prbplug,psfig]{article}

\begin{document}

\title{ {\small PHYSICAL REVIEW B64, 184509 (2001)}\\
\vspace{0.7cm} 
Ground State of the Three-Band Hubbard Model
}

\author{Takashi Yanagisawa, Soh Koike and Kunihiko Yamaji}

\address{Condensed-Matter Physics Group, 
Nanoelectronics Research Institute (NeRI), 
National Institute of Advanced Industrial Science and Technology (AIST), 
Central 2 1-1-1 Umezono, Tsukuba,
Ibaraki 305-8568, Japan}

\date{(18 October 2001)}

\maketitle

\begin{abstract}
The ground state of the two-dimensional three-band Hubbard model in the oxide 
superconductors
is investigated by using the variational Monte Carlo method.
The Gutzwiller-projected BCS and SDW wave functions are employed in search for a
possible ground state with respect to dependences on electron density.
Antiferromagnetic correlations are considerably strong near half-filling.
It is shown that the $d$-wave state may exist away from half-filling for
both the hole and electron doping cases.
Overall structure of the phase diagram obtained by our calculations
qualitatively agrees with experimental indications.
The superconducting condensation energy is in reasonable agreement with the
experimental value obtained from specific heat and critical magnetic field
measurements for optimally doped samples.
The inhomogeneous SDW state is also examined near 1/8 doping.
Incommensurate magnetic structures become stable due to hole doping in the
underdoped region, where the transfer $t_{pp}$ between oxygen orbitals plays
an important role in determining a stable stripe structure.\\
\\
PACS numbers: 74.20.-z,74.25.Dw,71.10.Fd.
\end{abstract}

\section{Introduction}
In order to investigate the mechanism of superconductivity (SC) in cuprate
high-$T_c$ superconductors\cite{lt99}, we examine the ground state of the
two-dimensional three-band Hubbard model
for CuO$_2$ planes which are contained usually
in their crystal structures.
It is believed that the CuO$_2$ plane contains the essential
features of high-$T_c$ cuprates.\cite{eme87,tje89}  
It is not an easy task to clarify
the ground state properties of the 2D three-band Hubbard model because of strong
correlations among $d$ and $p$ electrons.
We must treat the strong correlations properly to understand the phase
diagram of the high-$T_c$ cuprates.  
The quantum variational Monte Carlo method (VMC) is a tool to investigate
the overall structure of phase diagram from weak to strong correlation regions.
In this paper we investigate possible ground states in the three-band Hubbard
model for CuO$_2$ plane by employing VMC. 

Superconductivity in the one-band Hubbard model has been studied
by numerical\cite{hir85,whi89,fur92,dag94,hus96,kur97,zha97,nak97,yam98,yam01}
and analytical\cite{sca86,shi88,bic89,pao94,mon94,dah95} calculations.
The three-band Hubbard model
has also been investigated with intensive efforts
recently\cite{hir89,sca91,dop90,dop92,kur96,gue98,hot94,tak97,kob98,koi00}.
The exact diagonalization computations for the three-band model in early stage 
of high-$T_c$ research supported a possibility of superconductivity by
showing that holes can bind in small systems\cite{oga88,ste89}.
It is also reported that the attractive interaction works for both the $d$-wave
and extended-$s$ wave channels based on finite temperature quantum Monte
Carlo (QMC) simulations\cite{sca91}.
It has been shown recently that one can predict finite $T_c$ for the three-band
Hubbard
model based on perturbative calculations such as generalized RPA 
treatments\cite{hot94,tak97,kob98}.
In perturbative treatments of the one-band and three-band Hubbard models,
the spin fluctuations induced by the on-site Coulomb interaction promote
anisotropic pairing correlations.
QMC evaluations with some constraints due to the 
fermion sign problem are against a possibility of 
superconductivity in the three-band Hubbard model\cite{gue98}.

In order to investigate the possibility and origin of superconductivity, 
the recent work by Kondo is important where
it has been shown that the $d$-wave state has lower energy than the normal state
for small $U$ by employing the perturbation theory in $U$ for the one-band
Hubbard model.\cite{kon00}
This indicates that the ground state is superconductive with $d$-wave symmetry
for small values of $U$.
We can expect that this also holds for the three-band model\cite{koi01}.
It is then natural to expect that the $d$-wave state is stable for
finite $U$ unless there occurs some ordering in the ground state.
Among several possible long-range orderings, antiferromagnetic one should be
examined because the state with antiferromagnetic ordering is considerably 
stable near half-filling.
In fact, according to VMC work for the one-band Hubbard model, 
the antiferromagnetic (AF) energy gain is larger than the SC energy gain
by almost two order of magnitude near half-filling.
Then the competition between SC and AF states 
is very severe for the SC state\cite{yam98,yam01}.  
The SC region for the one-band Hubbard model is considerably restricted
and a possibility of pure superconducting state is very small\cite{yam98}.
A similar feature has been obtained by
VMC evaluations for $U_d=\infty$ three-band Hubbard
model\cite{asa96} where antiferromagnetic
region extends up to 50 percent doping and the $d$-wave phase exists only
in the infinitesimally small region near the boundary of antiferromagnetic phase.
Thus VMC results performed recently are consistent with the 
constrained path QMC calculations\cite{gue98} in the sense that a possibility of
$d$-wave phase for the one-band Hubbard model and $U_d=\infty$ three-band
Hubbard model is small at present, although an attractive interaction works
for $d$-wave pairing.

We expect that the antiferromagnetic region will shrink for the 
three-band Hubbard
model if we  adjust parameters contained in the model.
The parameters of the three-band Hubbard model are given by the Coulomb
repulsion $U_d$, energy levels of $p$ electrons $\epsilon_p$ and $d$
electron $\epsilon_d$, and transfer between $p$ orbitals given by $t_{pp}$.
A purpose of this paper is to investigate the property of antiferromagnetic
state and a competition between antiferromagnetism and superconductivity
for finite $U_d$ based on the three-band model following ansatz of 
Gutzwiller-projected wave functions.

It has also been argued that holes doped in the antiferromagnetically 
correlated 
spin systems induce incommensurate spin correlations in the ground state for
the one-band Hubbard model\cite{poi89,kat90,sch90,zaa96,sal96,ich99} and 
three-band model\cite{zaa89}  within the mean field approximation. 
In the mean-field treatment the energy scales appear to be extremely large
compared to values for real materials.
Recent neutron-scattering experiments revealed incommensurate spin
structures\cite{tra96,tra97,suz98,yama98,ara99,wak00,mat00,moo00} 
developed at low temperatures and at low energies.
The static incommensurate structure was reported on LSCO samples:
La$_{2-x}$Sr$_x$CuO$_4$, La$_{1.6-x}$Nd$_{0.4}$CuO$_4$ and
La$_{2-x}$Sr$_x$NiO$_{4+y}$.
The incommensurate magnetic peaks have been also reported for 
YBa$_2$Cu$_3$O$_{7-\delta}$ by the inelastic neutron-scattering experiments.
This type of inhomogeneous state may possibly provide a key concept to resolve
the anomalous properties of high-$T_c$ cuprates in the underdoped region.
We will examine a possible phase of incommensurate states
for the three-band Hubbard model by variational Monte Carlo method.

The paper is organized as follows.
In the next section the wave functions are presented.  The SC state and
uniform SDW state are discussed in Section III and a stability of 
incommensurate state is examined in the subsequent section.  A summary is given
in the last section.


\section{Hamiltonian and Wave Functions}

The Hamiltonian is given as\cite{gue98,asa96,yan00}
\onecolumn
\begin{eqnarray}
H&=& \epsilon_d\sum_{i\sigma}d^{\dag}_{i\sigma}d_{i\sigma} 
+ \epsilon_p\sum_{i\sigma}(p^{\dag}_{i+\hat{x}/2,\sigma}p_{i+\hat{x}/2,\sigma}
+p^{\dag}_{i+\hat{y}/2,\sigma}p_{i+\hat{y}/2,\sigma})\nonumber\\
&+& t_{dp}\sum_{i\sigma}[d^{\dag}_{i\sigma}(p_{i+\hat{x}/2,\sigma}
+p_{i+\hat{y}/2,\sigma}-p_{i-\hat{x}/2,\sigma}
- p_{i-\hat{y}/2,\sigma})+h.c.]\nonumber\\
&+& t_{pp}\sum_{i\sigma}[p^{\dag}_{i+\hat{y}/2,\sigma}p_{i+\hat{x}/2,\sigma}
-p^{\dag}_{i+\hat{y}/2,\sigma}p_{i-\hat{x}/2,\sigma}
-p^{\dag}_{i-\hat{y}/2,\sigma}p_{i+\hat{x}/2,\sigma}
+p^{\dag}_{i-\hat{y}/2,\sigma}p_{i-\hat{x}/2,\sigma} +h.c.]\nonumber\\
&+&U_d\sum_id^{\dag}_{i\uparrow}d_{i\uparrow}d^{\dag}_{i\downarrow}d_{i\downarrow}\nonumber\\
&=& H_0+V ,
\end{eqnarray}
\twocolumn
where
\begin{equation}
V = 
U_d\sum_id^{\dag}_{i\uparrow}d_{i\uparrow}d^{\dag}_{i\downarrow}d_{i\downarrow}.
\end{equation}
$\hat{x}$ and $\hat{y}$ represent unit vectors along $x$ and $y$ directions,
respectively.
$p^{\dag}_{i\pm\hat{x}/2,\sigma}$
and $p_{i\pm\hat{x}/2,\sigma}$ denote the operators for the $p$ electrons at
site $R_i\pm\hat{x}/2$.  Similarly $p^{\dag}_{i\pm\hat{y}/2,\sigma}$ and
$p_{i\pm\hat{y}/2,\sigma}$ are  defined.
Other notations are standard and energies are measured in units of
$t_{dp}$.
For simplicity we neglect the Coulomb interaction among $p$ electrons.

We consider the normal state, BCS and SDW wave functions with the Gutzwiller
projection.
These types of functions are standard wave functions and well describe the
ground-state properties with several long-range orderings.  They have
been investigated intensively for the one-band Hubbard 
model\cite{nak97,yam98,yam01,yan99,gro89,gia91,yam00,yan98}.  
In Refs.\cite{yan99,yan98} it has been discussed that they can be improved
systematically by operating correlation factors 
${\rm e}^{-\lambda H_0}{\rm e}^{-\alpha V}$.
For the model shown above they are written as
\begin{equation}
\psi_n= P_G\prod_{|k|\leq k_F,\sigma}\alpha_{k\sigma}^{\dag}|0\rangle ,
\end{equation}
\begin{equation}
\psi_{SC}= P_GP_{N_e}\prod_{k}(u_k+v_k\alpha^{\dag}_{k\uparrow}\alpha^{\dag}_{-k\downarrow})|0\rangle ,
\label{sc}
\end{equation}
\begin{equation}
\psi_{SDW}= P_G\prod_{|k|\leq k_F,\sigma}\beta_{k\sigma}^{\dag}|0\rangle ,
\end{equation}
where $\alpha_{k\sigma}$ is the linear combination of $d_{k\sigma}$,
$p_{xk\sigma}$ and $p_{yk\sigma}$ constructed to express an operator for the 
lowest band of a non-interacting Hamiltonian in the hole picture.  For $t_{pp}=0$,
$\alpha_{k\sigma}$ is expressed in terms of a variational parameter
$\tilde{\epsilon_p}-\tilde{\epsilon_d}$:
\onecolumn
\begin{equation}
\alpha^{\dag}_{k\sigma}= \left[\frac{1}{2}\left(
1+\frac{\tilde{\epsilon_p}-\tilde{\epsilon_d}}{2E_k}\right)\right]^{1/2}d^{\dag}_{k\sigma}
+{\rm i}\left[\frac{1}{2}\left(
1-\frac{\tilde{\epsilon_p}-\tilde{\epsilon_d}}{2E_k}\right)\right]^{1/2}\left(
\frac{w_{xk}}{w_k}p^{\dag}_{xk\sigma}
+\frac{w_{yk}}{w_k}p^{\dag}_{yk\sigma}\right),
\end{equation}
\twocolumn
where $w_{xk}=2t_{dp}{\rm sin}(k_x/2)$, $w_{yk}=2t_{dp}{\rm sin}(k_y/2)$,
$w_k=(w_{xk}^2+w_{yk}^2)^{1/2}$ and 
$E_k=[(\tilde{\epsilon_p}-\tilde{\epsilon_d})^2/4+w_k^2]^{1/2}$.
The Fourier transforms of $d$- and $p$- electron operators are defined as
\begin{equation}
d^{\dag}_{k\sigma}= \frac{1}{N^{1/2}}\sum_i d^{\dag}_{i\sigma}
{\rm e}^{{\rm i}k\cdot R_i} ,
\end{equation}
\begin{equation}
p^{\dag}_{xk\sigma}= \frac{1}{N^{1/2}}\sum_i p^{\dag}_{i+\hat{x}/2\sigma}
{\rm e}^{{\rm i}k\cdot (R_i+\hat{x}/2}) ,
\end{equation}
\begin{equation}
p^{\dag}_{yk\sigma}= \frac{1}{N^{1/2}}\sum_i p^{\dag}_{i+\hat{y}/2\sigma}
{\rm e}^{{\rm i}k\cdot (R_i+\hat{y}/2}) ,
\end{equation}
where $N$ is the total number of cells which consist of $d$, $p_x$ and $p_y$
orbitals.
Coefficients $u_k$ and $v_k$, appearing only as a ratio, are given by the BCS 
form:
\begin{equation}
\frac{v_k}{u_k} = \frac{\Delta_k}{\xi_k+(\xi_k^2+\Delta_k^2)^{1/2}} ,
\end{equation}
for $\xi_k=\epsilon_k-\mu$ where $\epsilon_k$ is the energy dispersion for
the lowest band.
$P_G$ is the Gutzwiller projection operator for the Cu $d$ site and $P_{N_e}$ 
is a projection
operator which extracts only the states with a fixed total electron number.
The SC order parameter $\Delta_k$ is assumed to have the following
$d_{x^2-y^2}$- and extended $s$-wave form:
\begin{equation}
{\rm d}~~~   \Delta_k= \Delta_s ({\rm cos}k_x-{\rm cos}k_y),
\end{equation}
\begin{equation}
{\rm s^*}~~~   \Delta_k= \Delta_s ({\rm cos}k_x+{\rm cos}k_y).
\end{equation}
Equation (\ref{sc}) is written as
\begin{equation}
\psi_{SC}= P_G \left( \sum_k \frac{v_k}{u_k}
\alpha^{\dag}_{k\uparrow}\alpha^{\dag}_{-k\downarrow}\right)^{N_e/2} .
\label{nbcs}
\end{equation}
The wave function given by eq.(13) agrees with
\begin{equation}
\psi_{BCS}= P_G \prod_{k}(u_k+v_k\alpha^{\dag}_{k\uparrow}\alpha^{\dag}_{-k\downarrow})|0\rangle ,
\label{bcs}
\end{equation}
in the thermodynamic limit. 
For the commensurate SDW state $\beta_{k\sigma}$ is given by a linear
combination of two wave numbers ${\bf k}$ and ${\bf k}+{\bf Q}$ for the 
commensurate vector ${\bf Q}=(\pi,\pi)$.
We can also investigate the incommensurate SDW state with incommensurate
vector ${\bf Q}= (\pi\pm 2\pi\delta,\pi)$ by diagonalizing the Hartree-Fock
Hamiltonian with antiferromagnetic long-range order.
The system sizes are given by $6\times 6$ and $8\times 8$ for the projected
BCS wave function and $16\times 4$, $24\times 6$, $32\times 8$,
$40\times 10$ and $16\times 16$ for the incommensurate SDW states. 
Our calculations are performed with the periodic and the antiperiodic
boundary conditions for the $x$- and $y$-direction, respectively.  This set
of boundary conditions was chosen so that $\Delta_k$ does not vanish for
any ${\bf k}$-points possibly occupied by electrons.

The expectation values are calculated following the standard Monte Carlo
procedure by using the Metropolis algorithm.
In the process of finding a minimum of energy, we should optimize many 
parameters included in the wave functions.  For such purpose we employ
correlated measurements method to reduce the required cpu time\cite{umr88}.

\begin{figure}
\centerline{\psfig{figure=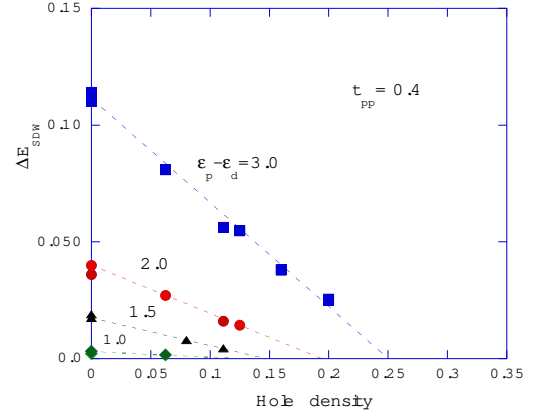,width=\columnwidth}}
\caption{
Energy per site $(E_{normal}-E)/N$ of the SDW state as a function of hole 
density $\delta$ for $t_{pp}=0.4$ and $U_d=8$.
From the top, $\epsilon_p-\epsilon_d=3$, 2, 1.5 and 1.
The results are for $6\times 6$, $8\times 8$, $10\times 10$ and $16\times 12$
systems.  Antiperiodic and periodic boundary conditions are imposed in $x$-
and $y$-direction, respectively.
Monte Carlo statistical errors are smaller than the size of symbols.
Curves are guide for eyes.
}
\label{aftp04}
\end{figure}

\begin{figure}
\centerline{\psfig{figure=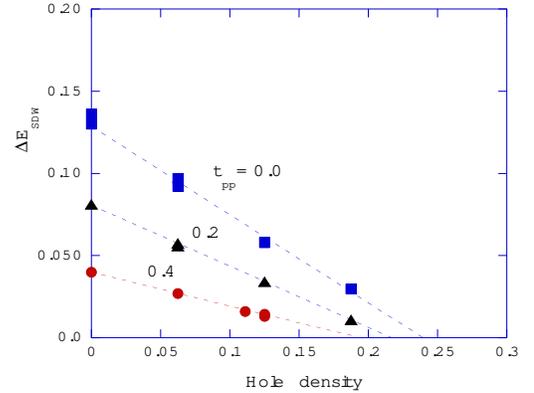,width=\columnwidth}}
\caption{
Energy per site $(E_{normal}-E)/N$ of the SDW state as a function of hole 
density $\delta$ for $t_{pp}=0.0, 0.2$ and 0.4 where $\epsilon_p-\epsilon_d=2$
and $U_d=8$.
The results are for $6\times 6$, $8\times 8$, $10\times 10$ and $16\times 12$
systems.
Curves are guide for eyes.
}
\label{afphas}
\end{figure}


\section{Condensation energy and phase diagram}
First, let us discuss the SDW phase near half-filling by evaluating
the ground-state energy for optimized parameters $g$,
$\tilde{\epsilon_p}-\tilde{\epsilon_d}$ and AF order parameter $\Delta_{AF}$.
We set $\epsilon_p=0$ throughout this paper.
It is expected that holes introduced by doping are responsible for the
disappearance of long-range antiferromagnetic ordering\cite{pre88,inu88,yan92}.
We show the SDW energy gain $\Delta E_{SDW}$ in Fig.1 as a function of doping 
ratio for several values of $\epsilon_p-\epsilon_d$.
$\Delta E_{SDW}$ increases and the SDW region becomes large  as 
$\epsilon_p-\epsilon_d$ increases.
The figure 2 shows the SDW energy gain for several values of $t_{pp}$,
where $\Delta E_{SDW}$ is reduced as $t_{pp}$ increases.
In Figs.3(a) and 3(b) the dependence on Coulomb repulsion $U_d$ is shown;
the SDW phase extends up to 30 percent doping when $U_d$ is large. 
Then it follows that the SDW region will be reduced if $\epsilon_p-\epsilon_d$ and $U_d$ 
decrease or $t_{pp}$ increases.  In fact,
Fig.4 shows the boundary of SDW phase in the $t_{pp}$-$\delta$ plane
for $U_d=8$ where $\delta$ is the hole density and negative density indicates
electron doping.
Compared to the calculations for $U_d=\infty$ the SDW region is reduced
greatly\cite{asa96}.

\begin{figure}
\centerline{\psfig{figure=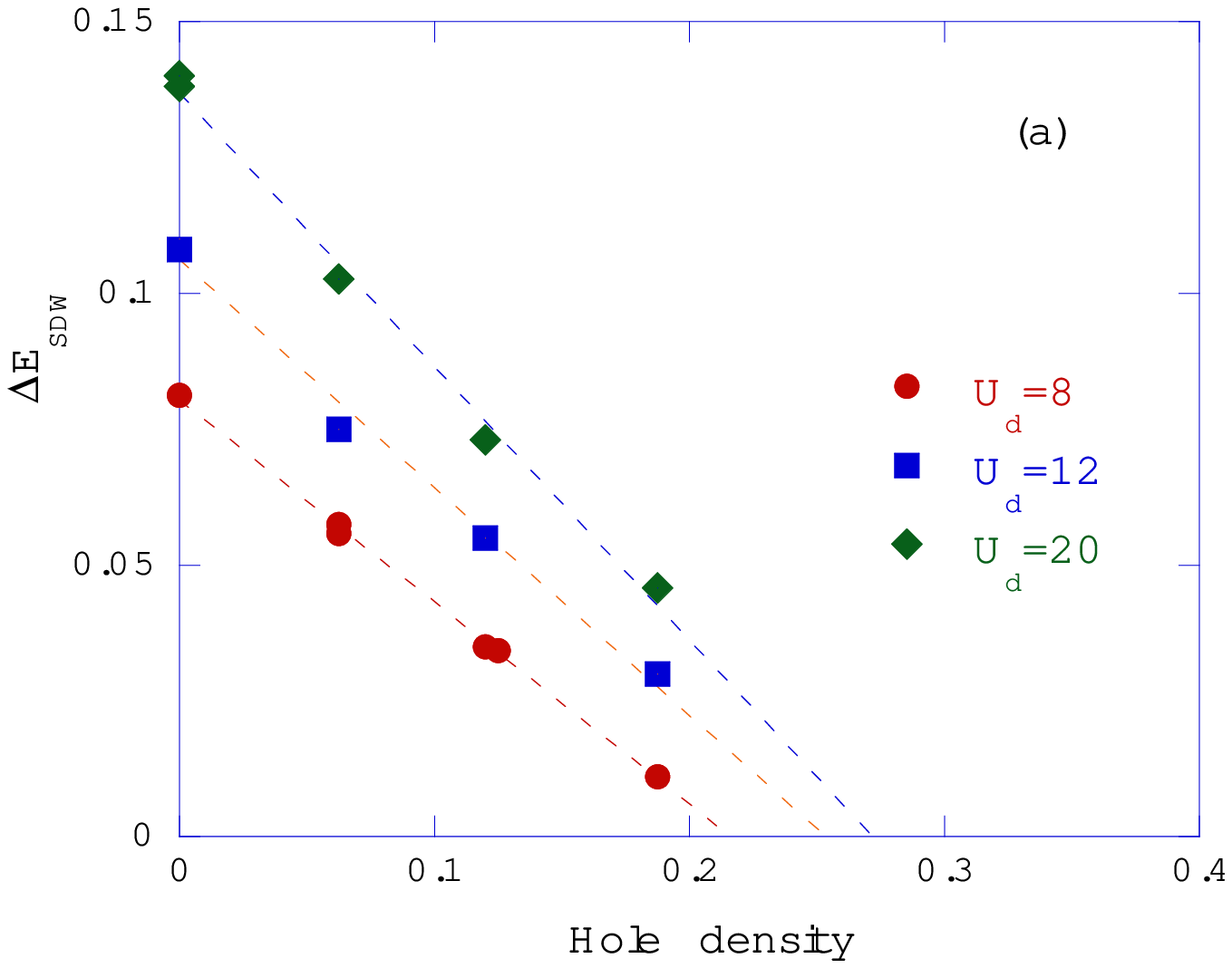,width=8cm}}
\centerline{\psfig{figure=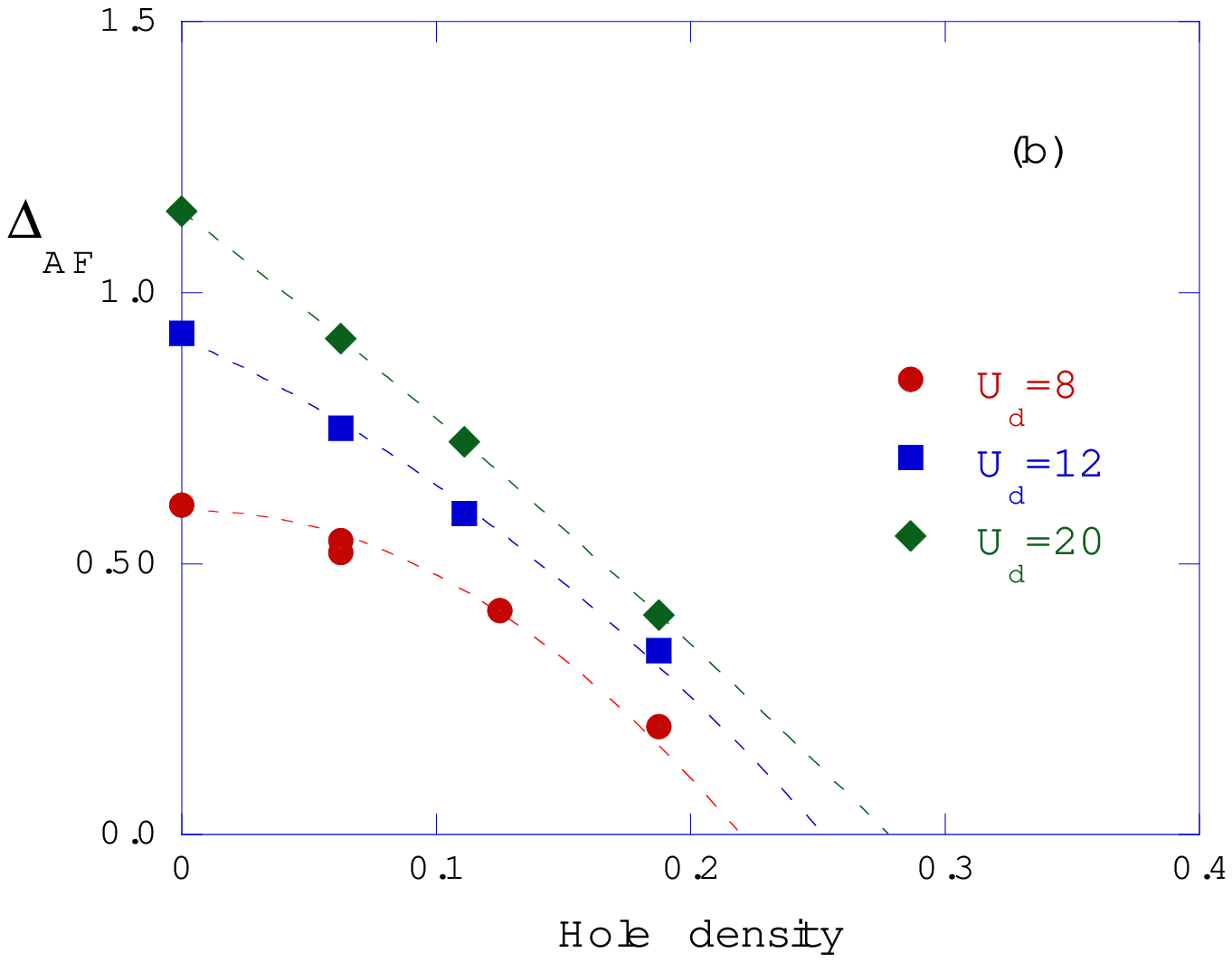,width=8cm}}
\caption{
(a) Energy per site $(E_{normal}-E)/N$ of the SDW state as a function of hole 
density $\delta$ for $U_d=8$, 12 and 20 where $\epsilon_p-\epsilon_d=2$ and 
$t_{pp}=0.2$.
(b) Antiferromagnetic order parameter as a function of hole density for 
$U_d=8$, 12 and 20 where $\epsilon_p-\epsilon_d=2$
and $t_{pp}=0.2$.
The results are for $6\times 6$, $8\times 8$, $10\times 10$ and $16\times 12$
systems.
Curves are guide for eyes.
}
\label{afu}
\end{figure}

\begin{figure}
\centerline{\psfig{figure=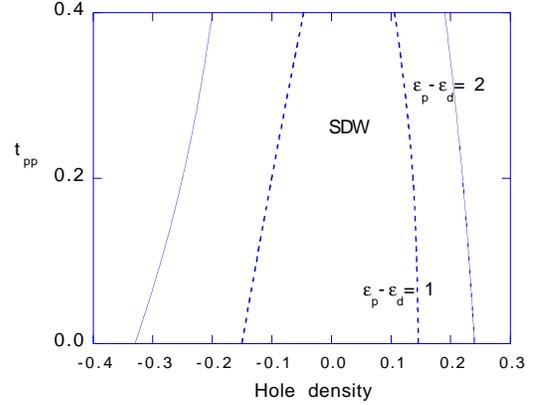,width=7cm}}
\caption{
Boundary of the SDW state in the plane of $t_{pp}$-$\delta$
for $\epsilon_p-\epsilon_d=2$ and 1.  We set $U_d=8$.}
\label{tpph}
\end{figure}

Next, let us turn to the projected-BCS wave function, where the
Gutzwiller parameter $g$, effective level difference 
$\tilde{\epsilon_p}-\tilde{\epsilon_d}$,
chemical potential $\mu$ and
superconducting order parameter $\Delta_s$ are considered as variational
parameters.  In Fig.5 we show the energy as a function of $\Delta_s$ where
$t_{pp}=0.0$, $U_d=8$ and $\epsilon_p-\epsilon_d=2$ and doping ratio is 
given by $\delta=0.111$
for (a) and $\delta=0.333$ for (b).
The $d$-wave superconductivity is most stable among various possible states 
such as isotropic $s$-wave and anisotropic $s$-wave pairing states.
The squares in Fig.5 denote the values for the normal state,
which are estimated independently by using an alternative Monte Carlo algorithm.
The finite SC energy gain indicates that the attractive interaction works for
$d$-wave pairing.

\begin{figure}
\centerline{\psfig{figure=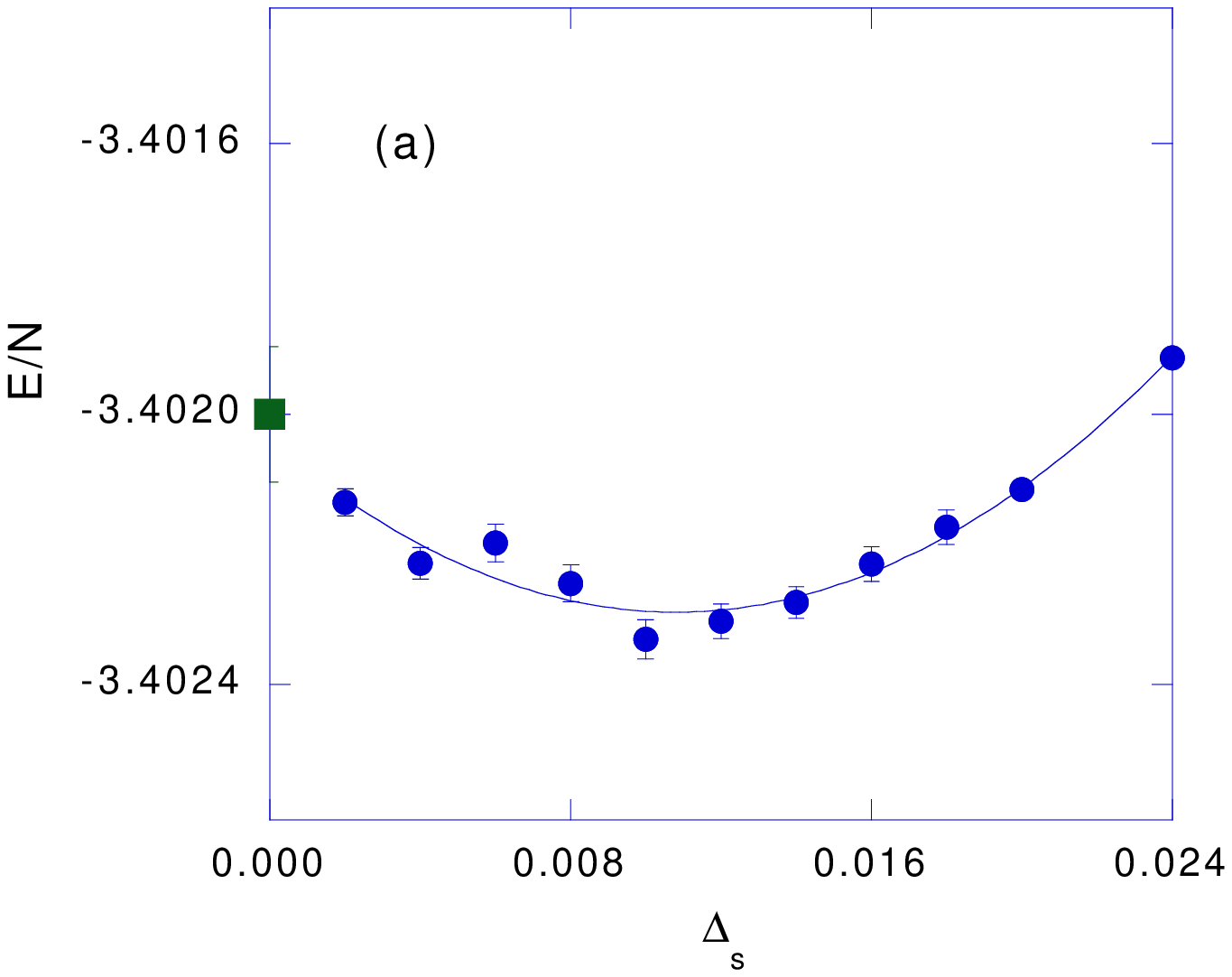,width=\columnwidth}}
\centerline{\psfig{figure=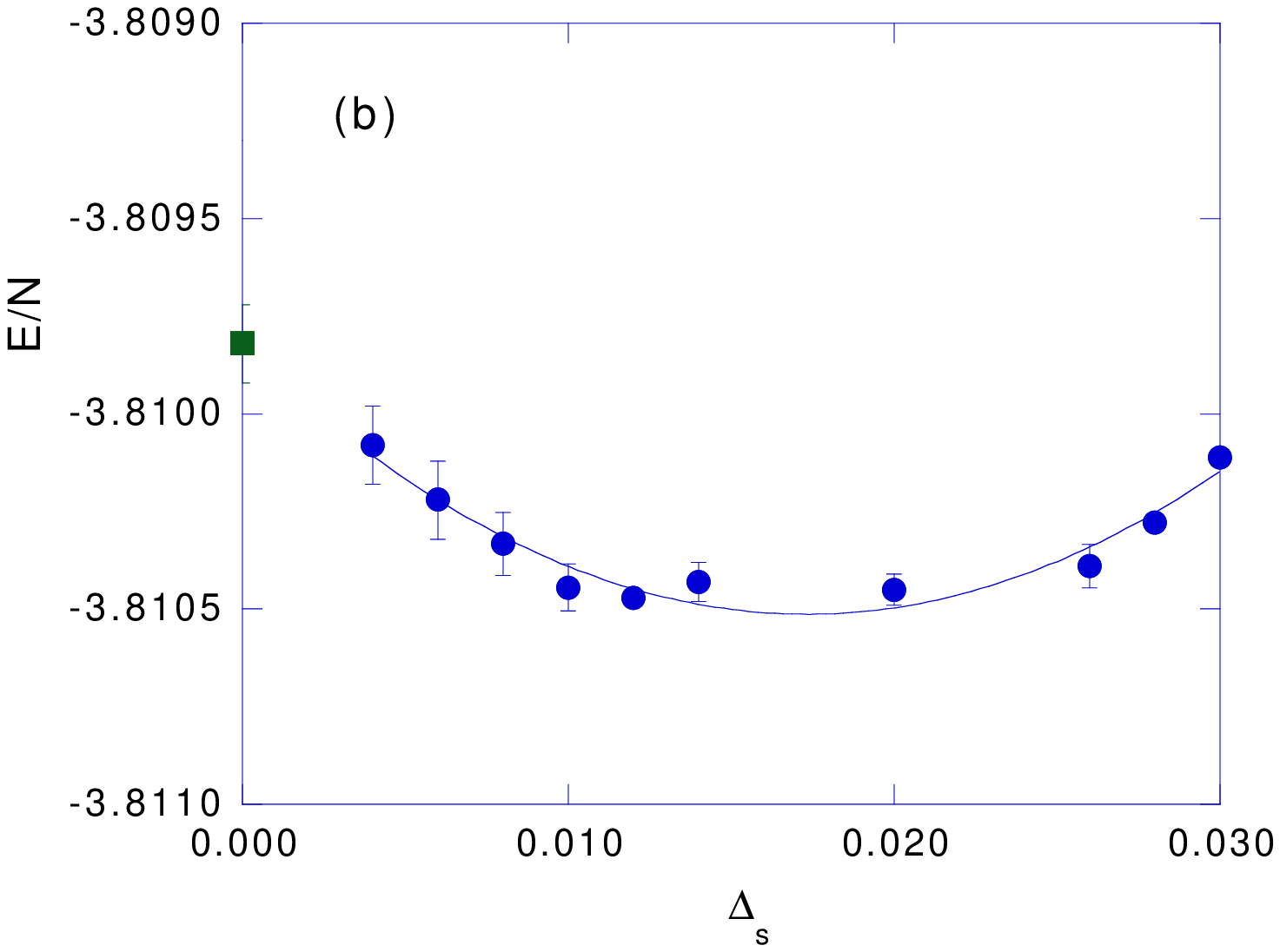,width=\columnwidth}}
\caption{
Ground state energy per site as a function of $\Delta_s$ on $6\times 6$
lattice for (a) $\delta=0.111$ and $t_{pp}=0.0$, and (b) $\delta=0.333$ and 
$t_{pp}=0.0$.  Parameters are given by
$U_d=8$ and $\epsilon_p-\epsilon_d=2$ in units of $t_{dp}$.
Squares denote the energies for the normal state
evaluated independently.
}
\label{energy}
\end{figure}

The SC energy gain (which is called the SC condensation energy in this paper)
is also dependent on $\epsilon_p-\epsilon_d$, as is
shown in Fig.6 for $t_{pp}=0.2$, $U_d=8$ and $\delta=0.111$ on $6\times 6$ 
lattice.  This shows a  
tendency that the SC condensation energy increases as $\epsilon_p-\epsilon_d$
increases, which is consistent with calculations for $U_d=\infty$.\cite{asa96}
It is noted that the dependence on $\epsilon_p-\epsilon_d$ for the SC
energy gain is rather weak compared to the SDW energy gain.
We also note that the SC energy gains for $U_d=8$ are mostly of the same order
of those for $U_d=\infty$.\cite{asa96}
 
From the calculations for the SDW wave functions, we should set 
$\epsilon_p-\epsilon_d$ and $U_d$ small so that the SDW phase does not
occupy a huge region near half-filling.  In Fig.7 we show
energy gains for both the SDW and SC states for $U_d=8$, $t_{pp}=0.2$ and
$\epsilon_p-\epsilon_d=2$, where the negative delta 
indicates the electron-doping case.  Solid symbols indicate the results for
$8\times 8$ and open symbols for $6\times 6$.
For this set of parameters the SDW region extends up to 20 percent doping
and the pure $d$-wave phase exists outside of the SDW phase.
The $d$-wave phase may be possibly identified with superconducting phase in
 the overdoped region in the
high-$T_c$ superconductors.

\begin{figure}
\centerline{\psfig{figure=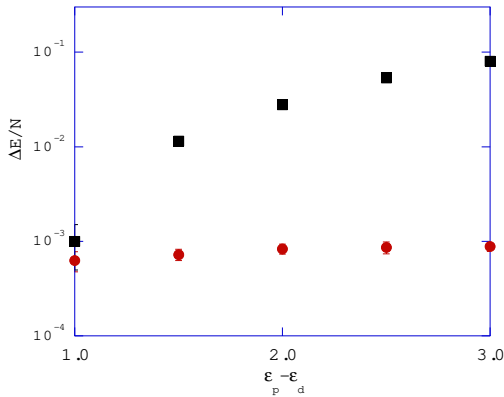,width=\columnwidth}}
\caption{
Superconducting (circles) and antiferromagnetic (squares) energy gains per 
site as a function of 
$\epsilon_p-\epsilon_d$ for $t_{pp}=0.2$ and $U_d=8$ on $6\times 6$ lattice.
}
\label{desc}
\end{figure}

\begin{figure}
\centerline{\psfig{figure=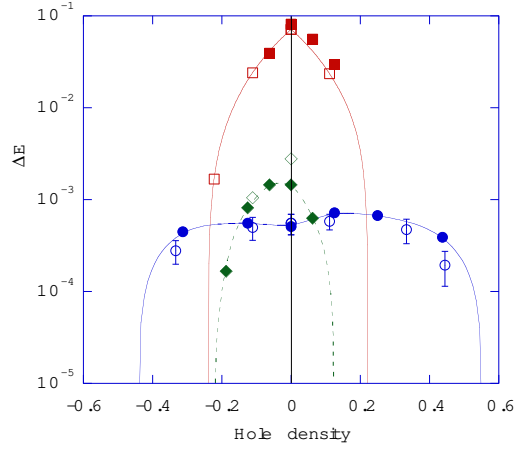,width=\columnwidth}}
\caption{
Condensation energy per site as a function of hole density $\delta$ for
$t_{pp}=0.2$, $\epsilon_p-\epsilon_d=2$ and $U_d=8$. 
Circles, squares and diamonds denote the energy gain per site in
reference to the normal state energy for $d$-wave, SDW and extended-$s$
wave states, respectively.
Solid symbols are for $8\times 8$ and open symbols are for $6\times 6$.
Curves are guide for eyes.
}
\label{phase}
\end{figure}

The superconducting condensation energy obtained by our calculations is
estimated as $E_{cond}\simeq 0.0005t_{dp}=0.75$meV per site in the overdoped
region near the boundary of SDW phase  
from the difference between the minimum and the intercept of the $E/N-\Delta_s$ 
curve with the vertical axis, where 
we set $t_{dp}=1.5$eV as estimated from cluster 
calculations\cite{esk89,hyb90,mcm90}.
We have also estimated $E_{cond}$ from several experiments such as specific
heat or critical field measurements for optimally doped samples.  
They are given as $0.17\simeq 0.26$meV from
specific heat data\cite{yam98,lor93,and98} and $0.26$meV from
critical magnetic field value $H^2_c/8\pi$\cite{yam98,hao91}. 
Our value is in reasonable agreement with the experimental data as was
already shown for the Hubbard model where the SC energy gain in the bulk
limit is given by 0.00117t/site=0.59meV/site\cite{yam00}.
This agreement between the theoretical and experimental condensation
energy is highly remarkable.
We expect that this value is not far from the correct
value according to the evaluations for improved wave functions\cite{yan99},
where it was shown that the energy gain is not changed so much due to
multiplicative correlation factors 
${\rm e}^{-\Delta\tau H_0}{\rm e}^{-\Delta\tau V}$.
We cannot estimate the SC condensation energy in the underdoped region
because the SDW state is more stable than $d$-wave state and the SC
condensation energy is not available experimentally due to a loss of
entropy in the underdoped region\cite{lor93}.

The phase structure obtained by our calculations agrees well with the
available phase diagram indicated by experiments qualitatively, which means
that a large SDW phase exists in the underdoped region and there is a 
$d$-wave superconducting phase next to SDW phase in the overdoped region.
Our calculations for electron-doping case predict $d$-wave symmetry away
from half-filling, which is consistent with recent experiments on
Nd$_{1.85}$Ce$_{0.15}$CuO$_{4-y}$\cite{tsu00}.

\begin{figure}
\centerline{\psfig{figure=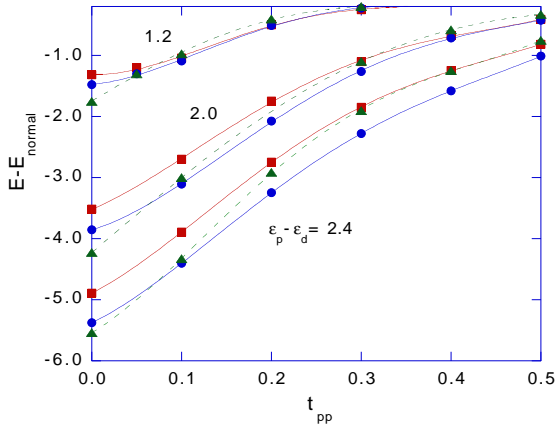,width=7.5cm}}
\caption{
Energies of commensurate and incommensurate SDW states on $16\times 4$ lattice
at $\delta=1/8$ for $U_d=8$.
Circles and triangles are for 4-lattice and 8-lattice stripes, respectively.
Squares denote energy for commensurate SDW state.
From the top $\epsilon_p-\epsilon_d=1.2$, 2.0 and 2.4.
We impose antiperiodic boundary condition in $x$-direction and periodic
boundary condition in $y$-direction.
Monte Carlo statistical errors are within the size of symbols.
}
\label{stripe1}
\end{figure}

\section{Incommensurate antiferromagnetism with spin modulation}

In this section let us discuss the underdoped region where the SDW state
is significantly stable as shown in the previous section.
Let us note that the SDW state can be possibly stabilized further if
we take into account a spin modulation in space, as has also been studied for
the one-band Hubbard model\cite{poi89,kat90,sch90,zaa96,sal96,ich99,gia90}
and the t-J model\cite{whi98,whi98b,hel99,kob99}.  
We can introduce a stripe in the uniform spin density state so that
doped holes occupy new levels close to the starting Fermi energy keeping 
the energy loss of antiferromagnetic background minimum.
The wave function with a stripe can be taken of the Gutzwiller type:
$\psi_{stripe}=P_G\psi_{stripe}^0$.
$\psi_{stripe}^0$ is the Slater determinant made from solutions of
the Hartree-Fock Hamiltonian\cite{gia90}
\begin{equation}
H_{stripe}=
H^0_{dp}+\frac{U}{2}\sum_{i\sigma}[\langle n_{di}\rangle
-\sigma(-1)^{x_i+y_i}\langle m_i\rangle]
d^{\dag}_{i\sigma}d_{i\sigma},
\end{equation}
where $H^0_{dp}$ is the non-interacting part of the Hamiltonian $H$ with
variational parameter $\tilde{\epsilon_p}-\tilde{\epsilon_d}$.
$\langle n_{di}\rangle$ and $\langle m_i\rangle$ are expressed in terms of 
modulation vectors
$Q_s$ and $Q_c$ for spin and charge part, respectively.
Including the constant part of $\langle n_{di}\rangle$ in the definition of
variational parameter $\tilde{\epsilon_d}$, we diagonalize the following
one-particle Hamiltonian to determine $\psi_{stripe}^0$:
\begin{equation}
H_{stripe}=
H^0_{dp}+\sum_{i\sigma}[\delta n_{di}
-\sigma(-1)^{x_i+y_i}m_i]
d^{\dag}_{i\sigma}d_{i\sigma}.
\end{equation}
$\delta n_{di}$ and $m_i$ are assumed to have the form 
\begin{equation}
\delta n_{di}=-\sum_j \alpha/{\rm cosh}((x_i-x_j^{str})/\xi_c),
\label{ndi}
\end{equation}
\begin{equation}
m_i=m\prod_j {\rm tanh}((x_i-x_j^{str})/\xi_s), 
\end{equation}
with parameters $\alpha$, $m$, $\xi_c$ and $\xi_s$.
$x_j^{str}$ denotes the position of a stripe.
In actual calculations we set $\xi_c=1$ and $\xi_s=1$ since
the energy expectation values are mostly independent of $\xi_c$ and $\xi_s$.
Since any eigenfunction of the Hamiltonian $H^0_{dp}$ can be a variational
wave function, we optimize $\alpha$ instead of fixing it in order to lower 
the energy expectation value further.
It is also possible to assume that $\delta n_{di}$ and $m_i$ oscillate
according to cosine curve given as cos($4\pi\delta x_i$) and 
cos($2\pi\delta x_i$), respectively.
Both methods give almost the same results within Monte Carlo statistical errors.

\begin{figure}
\centerline{\psfig{figure=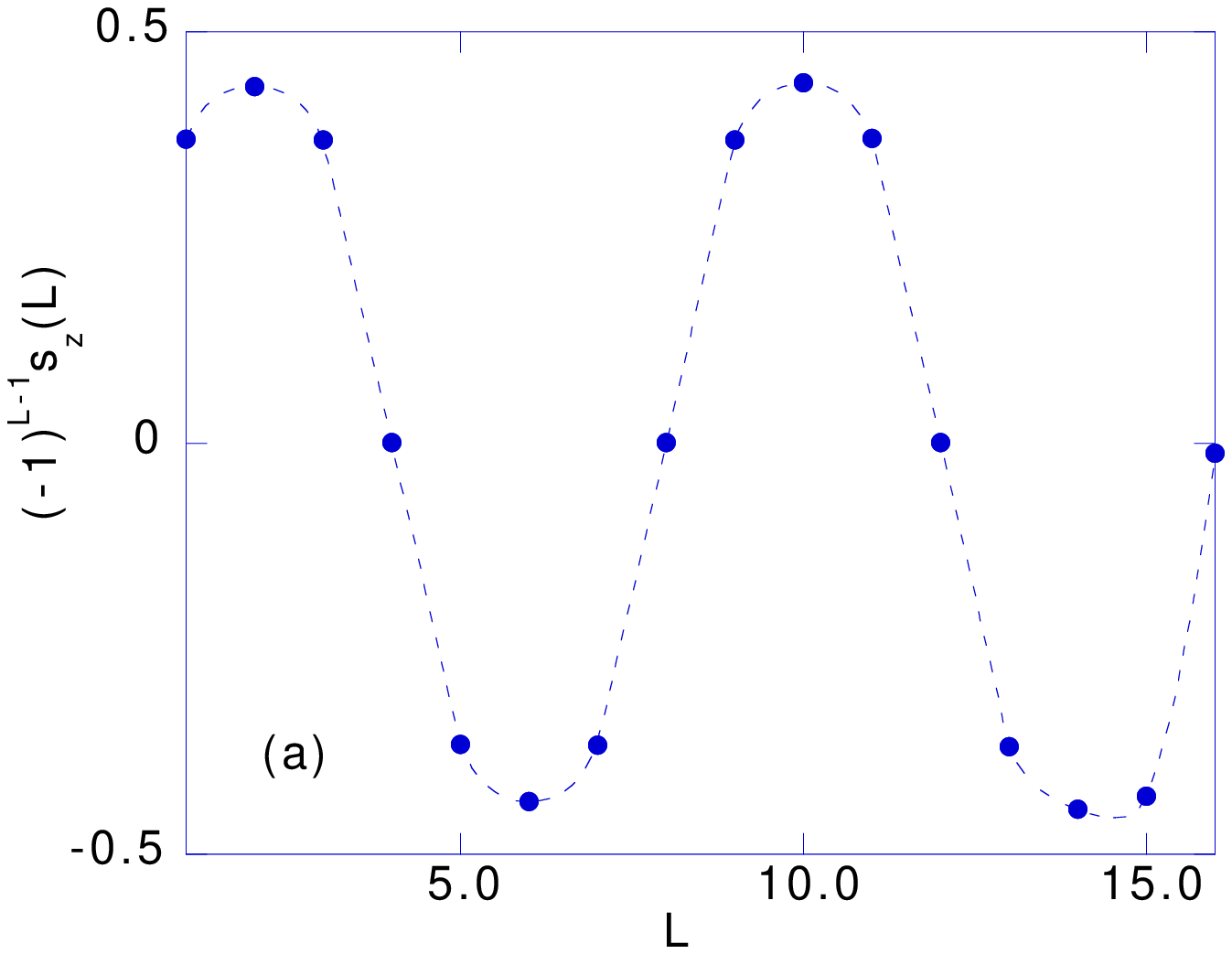,width=7cm}}
\centerline{\psfig{figure=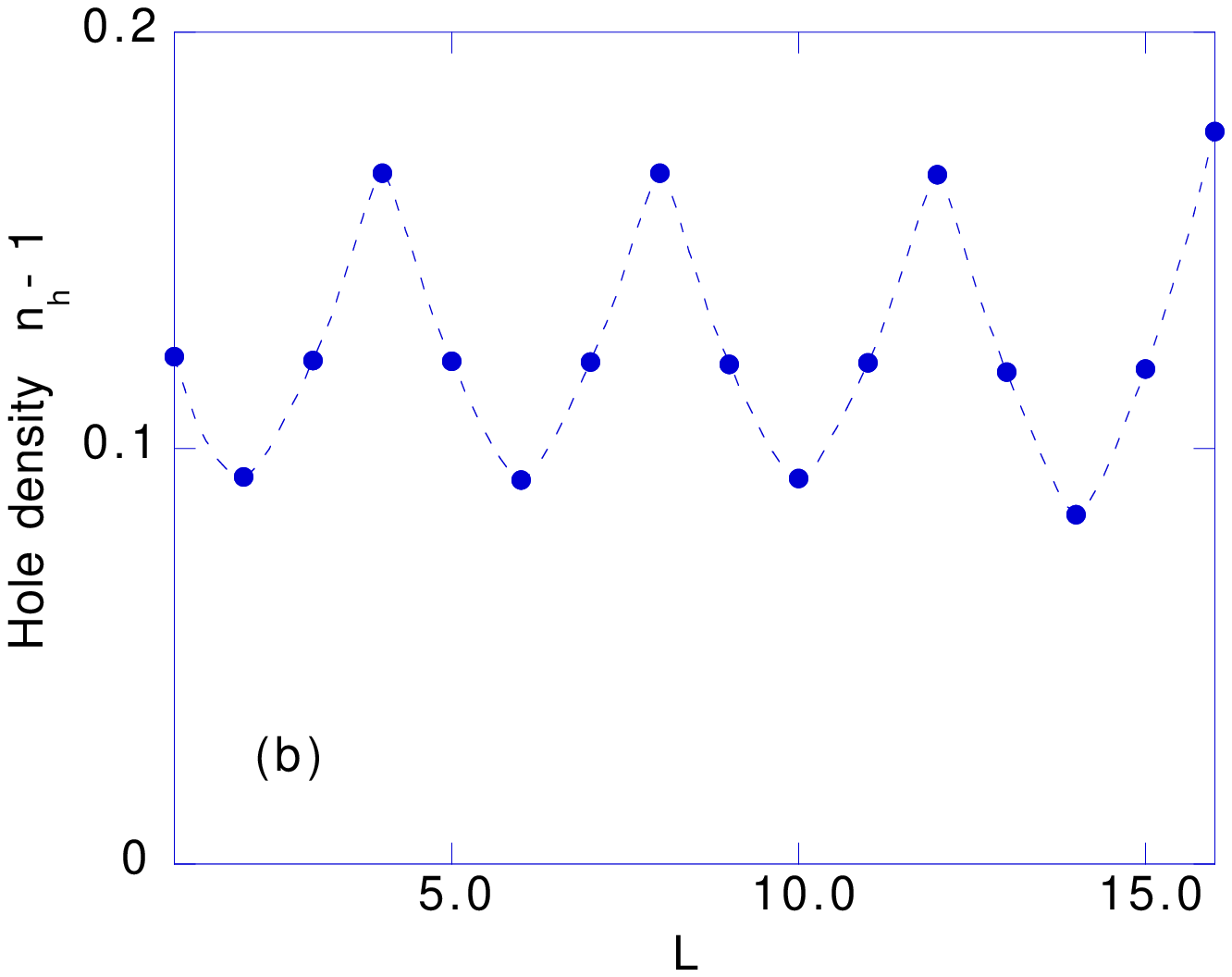,width=7cm}}
\caption{
Spin (a) and charge (b) densities for incommensurate state
at $\delta=1/8$ for $t_{pp}=0.4$, $U_d=8$ and $\epsilon_p-\epsilon_d=2$ on
$16\times 4$ lattice.  The boundary conditions are same as in Fig.8
}
\label{strspnd}
\end{figure}

\begin{figure}
\centerline{\psfig{figure=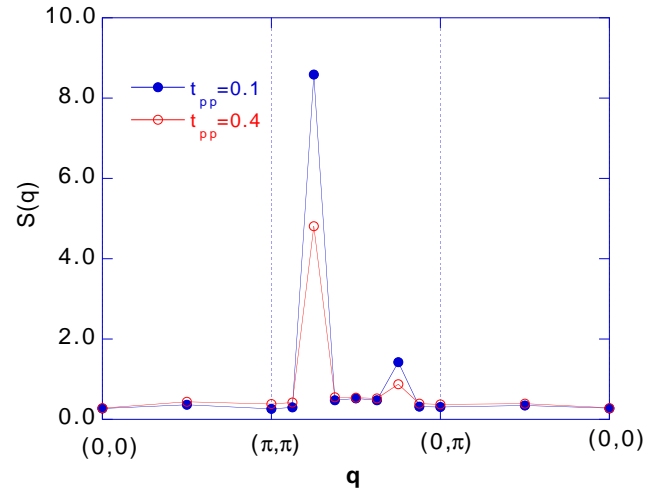,width=\columnwidth}}
\caption{
Spin structure function for incommensurate state
at $\delta=1/8$ for $t_{pp}=0.4$, 0.1 and $\epsilon_p-\epsilon_d=2$ on
$16\times 4$ lattice.  We set $U_d=8$.  The boundary conditions are same
as in Fig.8.
}
\label{szq}
\end{figure}

Recent neutron scattering experiments suggested that modulation vectors
are given by $Q_s=(\pi\pm 2\pi\delta,\pi)$ and $Q_c=(\pm 4\pi\delta,0)$
in the underdoped region, where $\delta$ denotes the doping ratio.
Here we define $n$-lattice stripe as an incommensurate state with one stripe per
n ladders for which $Q_s$ is given by $Q_s=(\pi\pm \pi/n,\pi)$.
Then the incommensurate state predicted by neutron experiments for 
$\delta=1/8$ is given by 4-lattice stripe.
For the three-band model, the transfer $t_{pp}$ between oxygen 
orbitals plays an important role to determine a possible SDW state.
If $t_{pp}$ is very large, the uniform SDW state is expected to be
stabilized because holes doped on oxygen sites can move around on the
lattice producing disorder effect on spin ordering uniformly.  
For small $t_{pp}$ the stripe states are considered to be 
realized\cite{zaa89}. 
Our motivation to consider non-uniform states for the three-band model lies in 
the idea that the distance between stripes may be dependent upon $t_{pp}$, i.e.
for small $t_{pp}$ the distance between stripes is large, for intermediate
values of $t_{pp}$ the 4-lattice stripe state is realized and for large $t_{pp}$
the uniform state or normal state is stable.

In Fig.8 we show the energies for commensurate and incommensurate SDW states
on $16\times 4$ lattice at $\delta=1/8$ as a function of $t_{pp}$, where
we impose the antiperiodic and periodic boundary conditions in $x$- and
$y$-direction, respectively, so that the closed shell structure is followed
for doped holes. 
We assumed that $\epsilon_p-\epsilon_d=1.2$, 2 and 2.4.
The 8-lattice stripe state for small $t_{pp}$ changes into uniform state
as $t_{pp}$ increases.  It shows that
incommensurate states become stable for large level difference
$\epsilon_p-\epsilon_d$.
The spin and charge densities of incommensurate state are shown in Fig.9
for $t_{pp}=0.4$ and $\epsilon_p-\epsilon_d=2$ where the charge density is
a sum of hole numbers on $d$-, $p_x$- and $p_y$-orbitals at site $L$.  
Spin density $S_z(i)=n_{di\uparrow}-n_{di\downarrow}$
vanishes at the positions of stripes associated with peaks of hole density.
The spin structure factor $S_z({\bf q})$ really has incommensurate peaks
as is shown in Fig.10.
The Figures 11(a) and 11(b) present the energies of incommensurate states
for $16\times 16$ lattice (which contains 768 atoms) where we set antiperiodic 
and periodic boundary
conditions in $x$- and $y$-direction, respectively, for (a) and in $y$- and $x$-
direction, respectively, for (b).  Both figures give almost the same results
as an evidence that the effect of boundary conditions is small for
$16\times 16$ system.  As expected, the structure of incommensurate state is
dependent upon the values of $t_{pp}$.

\begin{figure}
\centerline{\psfig{figure=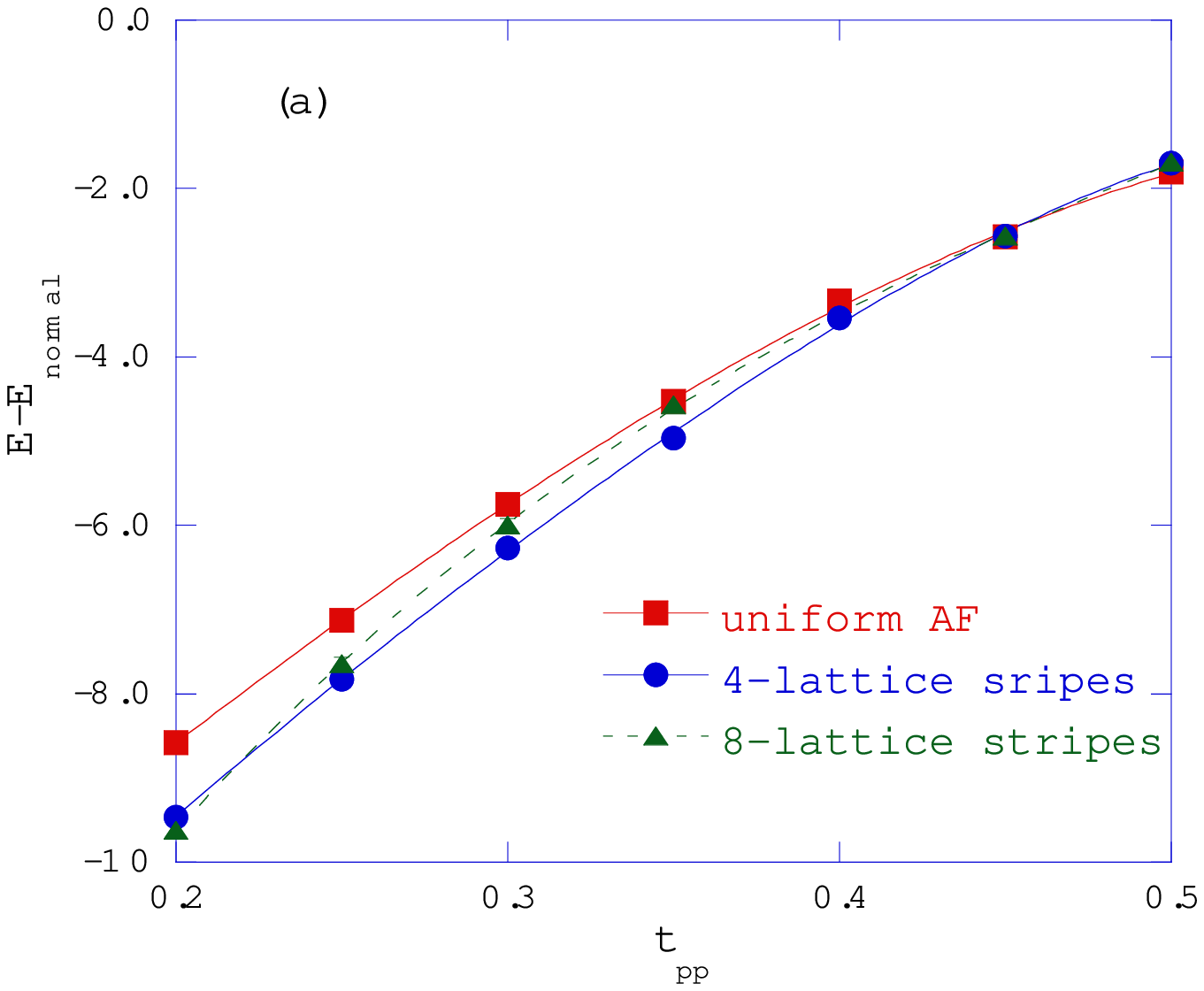,width=9cm}}
\centerline{\psfig{figure=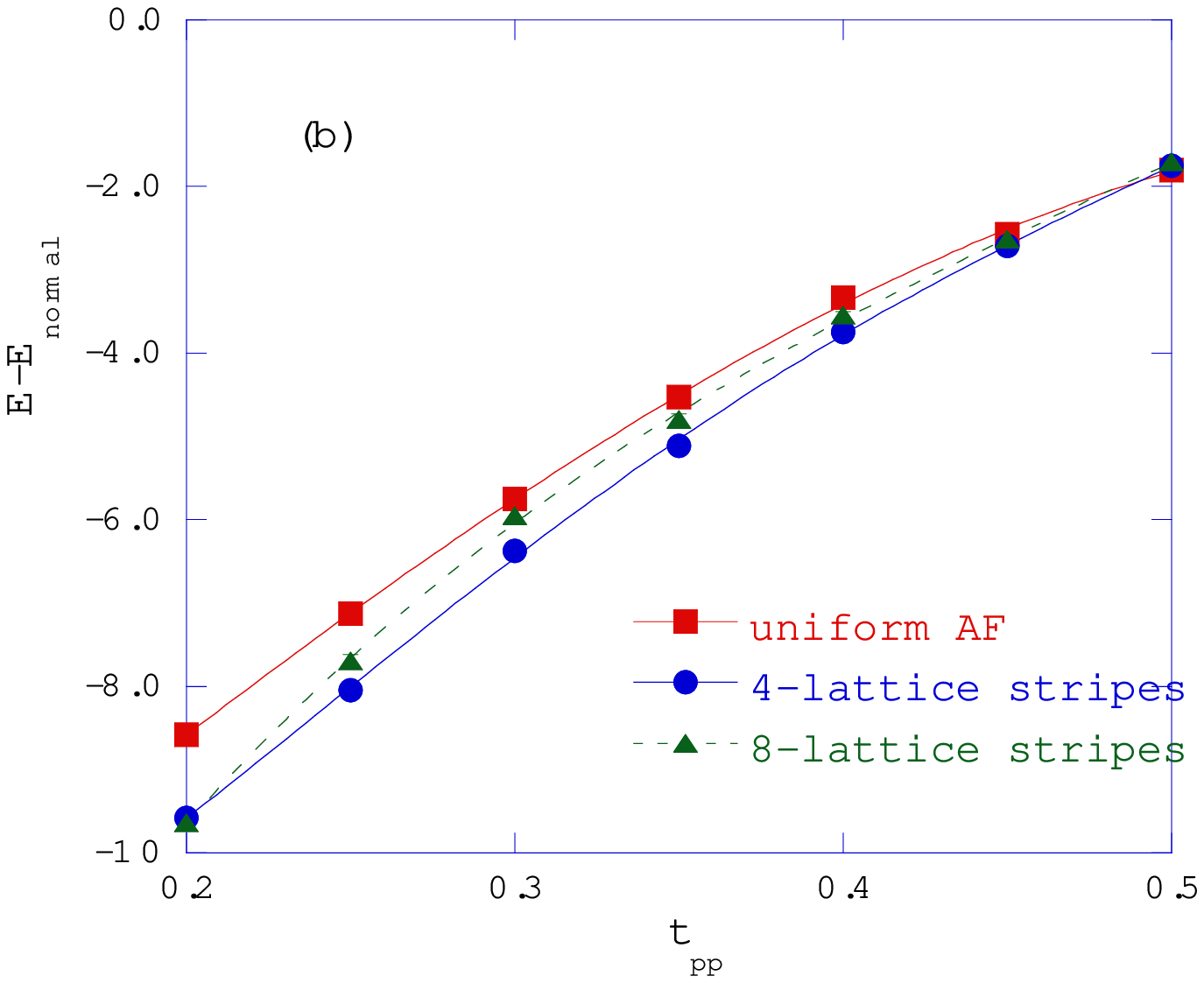,width=9cm}}
\caption{
Energies at $\delta=1/8$ for $16\times 16$ lattice.  
Parameters are given by $\epsilon_p-\epsilon_d=2$, $U_d=8$ and $t_{pp}=0.4$.
Symbols are the same as in Fig.8.
For (a) boundary conditions are antiperiodic and periodic in $x$- and 
$y$-direction,
respectively, and for (b) periodic and antiperiodic boundary conditions are
imposed in $x$- and $y$-direction, respectively.
Monte Carlo statistical errors are within the size of symbols.
}
\label{stripe3}
\end{figure}

Let us turn to a discussion of the energy gain due to a formation of
stripes, which is estimated from an extrapolation to the bulk limit as
shown in Fig.12.  One notes that the energy gain increases as the system
size increases.  
The energy gain per site for 4-lattice stripe state is
given by $\simeq 0.015t_{dp} \simeq 22.5$meV.  Furthermore the energy
difference between commensurate and incommensurate states is found to be
finite in the bulk limit, which is shown in Fig.13.  
Thus within VMC the stripe state with spin
modulation is stable at $\delta=1/8$ doping.

The antiferromagnetic order parameter $m$ in eq.(18) is of the 
order of
$0.5 t_{dp}\simeq 0.75$eV, 
while the SC order parameter $\Delta_s$ (which gives the minimum of energy) 
is of the order of
$0.01\sim 0.015 t_{dp}= 15$meV$\sim 20$meV at $\delta\sim 0.2$.   
The magnitude of SC order parameter agrees with
measurements of tunneling spectroscopy\cite{kas98,ido98} where $\Delta_s$ is
estimated as $\Delta_s\simeq 17$meV for YBCO sample.\cite{kas98}  
The antiferromagnetic order parameter is larger
than SC order parameter at least by one order of magnitude.
The charge order parameter $\alpha$ in eq.(17) is small and negligible
compared to $m_i$.

\begin{figure}
\centerline{\psfig{figure=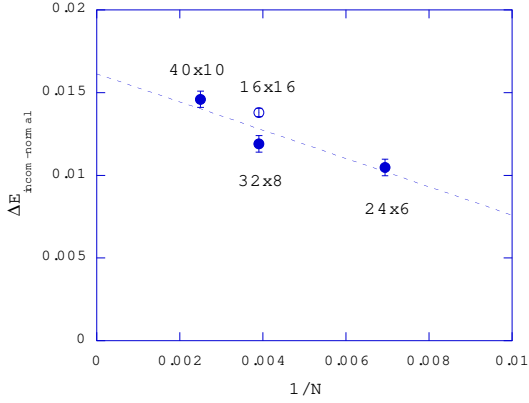,width=\columnwidth}}
\caption{
$E_{normal}-E_{incom}$ as a function of 1/N.
Parameters are given by $t_{pp}=0.4$, $U_d=8$ and $\epsilon_p-\epsilon_d=2$.
Solid circles are for $24\times 6$, $32\times 8$ and $40\times 10$.
Open circle is for $16\times 16$.
We set antiperiodic and periodic boundary conditions in $x$- and $y$-
direction, respectively.
}
\label{enestr}
\end{figure}

\begin{figure}
\centerline{\psfig{figure=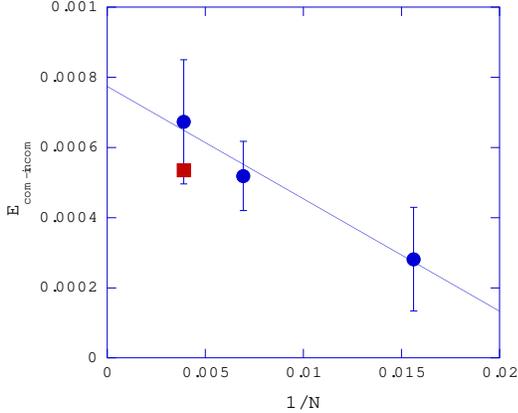,width=\columnwidth}}
\caption{
$E_{com}-E_{incom}$ as a function of 1/N.
Parameters are given by $t_{pp}=0.4$, $U_d=8$ and $\epsilon_p-\epsilon_d=2$.
Circles are for rectangular lattices and the square is for $16\times 16$.
Boundary conditions are the same as in Fig.12
}
\label{eneinc}
\end{figure}

\section{Summary}
We have presented our evaluations for the 2D three-band Hubbard model based on
the variational Monte Carlo method.
Our work is regarded as a starting step for more sophisticated calculations
in future such as the inclusion of correlation factors of Jastrow type or
Green function Monte Carlo approaches.
The SC energy scales obtained from our evaluations are consistent with
experimental indications, which provides a support to our approaches.

According to VMC the attractive interaction works for $d$-wave pairing due to
electron correlations.
The strength of $U_d$ is also important to determine the phase boundary
of the SDW phase.  If $U_d$ is extremely
large, the SDW region extends up to large doping for which the
$d$-wave region is restricted to infinitesimally small region near the
boundary of antiferromagnetic phase.
For intermediate values of $U_d$ and $\epsilon_p-\epsilon_d$ the SDW region 
is reduced and the $d$-wave superconducting phase may exist.
The fact that the SC condensation energy agrees reasonably with the
experimental data for optimally doped samples supports our computations.  
The magnitude of SC order
parameter is also consistent with tunneling spectroscopy experiments.  
From our data for $\Delta E_{SC}$ and $\Delta_s$ and the relation
$N(0)\Delta_s^2/2=\Delta E_{SC}$, the effective density of state $N(0)$ can be
estimated as $N(0)\simeq 3\sim 6.7({\rm eV})^{-1}\simeq 4.4\sim 10/t_{dp}$
at $\delta\sim 0.2$ in the overdoped region,
which is not far from the BCS estimate $N(0)\sim 2$ to 3(eV)$^{-1}$ by using
$N(0)(k_BT_c)^2/2$ for optimally doped YBCO.\cite{and98}
We expect that the pure $d$-wave state from optimal to overdoped regions
may be described by the projected-BCS wave function.
The phase diagram for electron-doping is consistent with the available
experimental indications suggesting that the properties of electron-doped  
materials may be understood within our approach.
In the SDW region the incommensurate spin structures are  stabilized for 
the low-doping case to keep the energy loss minimum due to disorder effect
caused by holes.
A competition among the uniform SDW state, SDW state with stripes,
and pure $d$-wave SC is highly non-trivial.
A picture for the hole-doping case followed from our evaluations is that a
stripe state is stable in the underdoped region and changes into the $d$-wave 
SC in the overdoped region.

We thank S. Koikegami for valuable discussions.  The computations were carried out
on the alpha cluster machines supported by the ACT-JST program of Japan Science and 
Technology Cooperation, and also on the HItachi SR8000 at the Tsukuba Advanced
Computing Center of the National Institute of Advanced Industrial Science and
Technology.

\end{document}